\documentclass[10pt,conference]{IEEEtran}
\IEEEoverridecommandlockouts
\newif{\ifshowcomments}
\showcommentsfalse

\usepackage{amsmath,amssymb,amsfonts}
\usepackage{placeins}
\usepackage[utf8]{inputenc}
\usepackage{graphicx}
\usepackage{epstopdf}
\usepackage{textcomp}
\usepackage{bm}
\usepackage[ruled,linesnumbered,vlined]{algorithm2e} % for pseudocode
\usepackage{tcolorbox}
\usepackage{braket}
\usepackage{amsthm}
\usepackage[hidelinks]{hyperref}
\usepackage{mathtools}
\usepackage{float}
\usepackage{booktabs} % Für professionelle Linien
\usepackage[inline]{enumitem}
\usepackage{caption}
\usepackage{tabularx}
\usepackage{siunitx} % Für perfekte Zahlausrichtung
\usepackage{colortbl}
\usepackage{xcolor}
\usepackage{xfp}

\SetAlgorithmName{Algorithm}{Algorithm}{List of Algorithms}

\newcommand{\vmin}{-0.06} % Für TVD
\newcommand{\vmax}{0.06} % Für TVD
\newcommand{\tmin}{-0.35} % Für Transmission
\newcommand{\tmax}{0.35} % Für Transmission
\definecolor{TUM_orange}{RGB}{255,0,0}
\definecolor{TUM_green}{RGB}{162,173,0}
\definecolor{my_green}{HTML}{32CD32}
\definecolor{my_red}{HTML}{D2042D}
\colorlet{positivecol}{my_green}
\colorlet{negativecol}{my_red}

\newcommand{\ptvd}[1]{%
    \ifnum\fpeval{#1 <= 0} = 1
        \edef\intensity{\fpeval{max(0, min(100, round(#1 / \vmin * 100, 0)))}}%
        \colorlet{valcol}{positivecol!\intensity!black}%
        \textcolor{valcol}{{$#1$}}%
    \else
        \edef\intensity{\fpeval{max(0, min(100, round(#1 / \vmax * 100, 0)))}}%
        \colorlet{valcol}{negativecol!\intensity!black}%
        \textcolor{valcol}{{$+#1$}}%
    \fi
}

\newcommand{\ptrans}[1]{%
    \ifnum\fpeval{#1 >= 0} = 1
        \edef\intensity{\fpeval{max(0, min(100, round(#1 / \tmax * 100, 0)))}}%
        \colorlet{valcol}{positivecol!\intensity!black}%
        \textcolor{valcol}{{$+#1$}}%
    \else
        \edef\intensity{\fpeval{max(0, min(100, round(#1 / \tmin * 100, 0)))}}%
        \colorlet{valcol}{negativecol!\intensity!black}%
        \textcolor{valcol}{{$#1$}}%
    \fi
}

\makeatletter
\let\MYcaption\@makecaption
\makeatother
\usepackage{subcaption} % provides the environment subfigure
\makeatletter
\let\@makecaption\MYcaption
\makeatother
\newtheorem{example}{Example}

\usepackage[
    style = ieee,
    citestyle = ieee-comp,
    maxnames=4,
    minnames=1,
    date=year,
    doi=false,
    isbn=false,
    backend=biber
]{biblatex}

\def\BibTeX{{\rm B\kern-.05em{\sc i\kern-.025em b}\kern-.08em
    T\kern-.1667em\lower.7ex\hbox{E}\kern-.125emX}}

\DeclareFieldFormat
  [misc, book]
  {title}{\mkbibquote{#1}}

\AtEveryBibitem{
  \clearname{editor}%
  \clearfield{series}%
  \clearfield{isbn}%
  \clearfield{issn}%
  \clearfield{volume}%
  \clearfield{number}%
  \clearfield{pages}%
  \clearfield{url}%
  \clearfield{doi}%
}

\ifshowcomments
\usepackage[textwidth=5cm]{todonotes} % <-- maybe change width of todonotes
\usepackage[switch]{lineno} % not compatible with package `flushend`
\linenumbers
\newlength{\widen}
\advance\paperwidth by \dimexpr\widen+\widen\relax
\advance\evensidemargin by \widen
\advance\oddsidemargin by \widen
\marginparwidth=\dimexpr\widen+0mm\relax % <-- maybe you need to modify the 0mm as well
\newcommand{\mytodo}[4]{\bgroup\color{#2!65!black}#3\egroup\todo[size=\scriptsize,color=#2]{#1: #4}}
\else
\newcommand{\mytodo}[4]{#3}
\usepackage{balance} % not compatible with package `lineno`
\fi
\definecolor{yannick}{HTML}{e89c1c}
\definecolor{robert}{HTML}{e5dd0c}
\definecolor{tobi}{HTML}{76c11f}
\definecolor{andi}{HTML}{f0d1ff}
\begin{document}

\title{Handling Imperfections in Subcircuit Compilation for Linear Optical Quantum Computing}

\author{
    \IEEEauthorblockN{Tobias Forster\IEEEauthorrefmark{1}\hfill
    Yannick Stade\IEEEauthorrefmark{1}\hfill
    Andreas Fyrillas\IEEEauthorrefmark{3}\hfill
    Jean Senellart\IEEEauthorrefmark{3}\hfill
    Lukas Burgholzer\IEEEauthorrefmark{1}\IEEEauthorrefmark{2}\hfill
    Robert Wille\IEEEauthorrefmark{1}\IEEEauthorrefmark{2}%
    }
    
    \IEEEauthorblockA{\IEEEauthorrefmark{1}Chair for Design Automation, Technical University of Munich, Munich, Germany}
    \IEEEauthorblockA{\IEEEauthorrefmark{2}MQSC, Garching near Munich, Germany}
    
    % Combined third affiliation and email table to save vertical "padding" space
    \IEEEauthorblockA{\IEEEauthorrefmark{3}Quandela, Massy, France\\[1ex]
        \centering
        \begin{tabular}{c@{\hskip 30pt}c}
            \{t.forster, yannick.stade, lukas.burgholzer, robert.wille\}@tum.de & 
            \{andreas.fyrillas, jean.senellart\}@quandela.com \\
            \href{https://www.cda.cit.tum.de/research/quantum}{www.cda.cit.tum.de/research/quantum} & 
            \href{https://www.quandela.com}{www.quandela.com}
        \end{tabular}
        \vspace{-5mm} % This is the "Magic Number" to fix the staggered column drop
    }
}
\vspace{-1mm}
\maketitle

\begin{abstract}
\emph{Linear Optical Quantum Computing}~(LOQC) is emerging as a promising technology in large-scale quantum computing.
Corresponding devices, including linear optical circuits, are rapidly growing in size and fabrication precision.
However, even with improved quality, the circuits still exhibit imperfections leading to photon loss or phase noise that vary across the circuit.
Already small deficiencies can have a vast impact on the results in both accuracy and efficiency.
For computations that do not require the entire linear optical circuit, so-called \emph{subcircuits}, one can choose where to perform the computation, opening up optimization opportunities that are addressed by \emph{subcircuit compilation}.
Current compilers spread the subcircuit's computation across the entire circuit, connecting the same input and output ports, thereby failing to fully exploit the optimization opportunities.
However, given the subcircuit's reduced size compared to the entire circuit, one can keep its implementation localized and intentionally route photons to eventually connect good input and output ports.
This, however, is a highly non-trivial task, as numerous options must be weighed against one another, and the quality of the solution is influenced by a wide range of hardware parameters.
In particular, the reflectivities of imperfect beam splitters can lead to photon losses during routing, which must be traded off against losses at the input and output ports due to imperfect transmission rates.
Since current compilers do not account for these effects, we propose the first efficient subcircuit compilation method that optimizes for the coincidence \mbox{rate---the} probability of successfully measuring all desired photons---by handling hardware imperfections.
Our evaluations demonstrate significant improvement over existing methods in robustness against phase noise and significantly increased coincidence rates across all benchmarks of up to plus 33.2\%.
\end{abstract}

\begin{IEEEkeywords}
    linear optical quantum computing, subcircuit compilation, hardware-aware optimization
\end{IEEEkeywords}
\vspace{-2mm}

\section{Introduction}

Photonic quantum computers perform computational tasks by harnessing specific properties of quantum states of light, such as single-photons. Single-photons are good candidates for quantum information carriers due to their low susceptibility to decoherence, compatibility with room-temperature operation, and the maturity of optical components. Current photonic platforms provide a promising route toward near-term computational~\cite{Aaronson2013, Deng2023} and energetic~\cite{Soret2026} quantum advantage.

A typical photonic device for \emph{Linear Optical Quantum Computing}~(LOQC~\cite{Kok2007}) that uses single-photons consists of three parts \cite{Qiang2018, Flamini2018, Wang2019, Vigliar2021, Chi2022, VanDerMeer2023, Maring2024, Skryabin2025}:
\begin{enumerate*}
    \item sources producing indistinguishable single-photons,
    \item these single-photons are routed towards a linear optical circuit such that multiple photons enter the circuit simultaneously in different input ports, and
    \item at the end of the circuit, photons are detected at each output port.
\end{enumerate*}
The circuit implements a linear optical transformation, which, together with the photon detection, constitutes the computation.
Such architectures can implement sampling tasks in which one records the simultaneous detection of $n$ photons.
Prominent examples include boson sampling~\cite{Brod2019}, photonic implementations of variational quantum algorithms~\cite{Cerezo2021}, quantum simulation~\cite{Aspuru-Guzik_Walther_2012}, and applications to graph problems~\cite{Mezher2023}.

The linear optical circuits are typically integrated devices that enable controlled multi-photon interference \cite{Harris2018, Wang2020}. They consist of waveguides that confine light along predefined paths, beam splitters that couple pairs of waveguides \cite{Dong2017} (enabling photons to be distributed between them), and reconfigurable phase shifters \cite{Harris2014} (which apply a phase shift to an individual waveguide). When arranged in mesh architectures such as those proposed by Reck~\cite{Reck1994}, Clements~\cite{Clements2016}, or Bell–Walmsley~\cite{Bell2021}, these circuits can implement any linear optical transformation by tuning the circuit's phase shifters.

Every linear optical transformation is described by a unitary matrix, which determines the performed computation on the photons.
Finding a set of phases with which the circuit applies the desired computation specified by a target unitary matrix is done by a compiler in the process of \emph{phase compilation}~\cite{Reck1994, Clements2016, Bell2021, Kumar2021, Bandyopadhyay2021, Mower2015, Pai2019, Fyrillas2024}.
In the scenario of the target unitary's dimension matching the number of waveguides of the physical circuit, this compilation's optimization potential is limited.
However, if the desired computation by the target unitary is smaller than the circuit, which we refer to as the \emph{subcircuit}, it can be placed arbitrarily on the circuit, offering new optimization opportunities to handle hardware imperfections and improve overall result quality.

One key limitation of LOQC experiments is the circuit's optical transmission rate, defined as the ratio of photons exiting the circuit to those injected.
Given the circuit imperfections, such as input and output ports that are highly susceptible to photon loss \cite{Son2018} (thus, low transmission rates), the overall transmission rate is limited.
This is particularly problematic for multi-photon experiments, as the probability of measuring all injected photons, the \emph{coincidence rate}, scales as $T^n$, with the transmission rate $ T$ and the number of photons $n$. 
Together with the fact that many LOQC computations are inherently probabilistic \cite{Ralph2002, Knill2002}, this highlights the critical importance of maximizing transmission rates for successful scaling.

Notably, the coincidence rate is heavily influenced by the individual transmission rates of the considered input and output ports, which can be determined by calibrating the circuit \cite{Fyrillas2024}.
However, current unitary-to-phase compilation protocols do not account for these hardware imperfections and potentially use the entire circuit to implement computations, even when the target unitary would require only a subset of the circuit's resources. 
Compared to keeping the computation zone compact, this larger computation zone results in lower coincidence rates and increased susceptibility to phase noise as more errors accumulate.

In this work, we propose a compilation strategy tailored to harness the optimization potential of compiling subcircuits and, by doing so, handling the imperfections of the circuit.
To this end, it places the subcircuit into the circuit to a specific location, referred to as the \emph{computation zone}, and actively routes photons to that zone.
Within this process, it balances the transmission rates of the input and output ports of the circuit with those induced by routing through imperfect beam splitters.
In simulated experiments, we demonstrate that this compilation strategy is less prone to phase noise and yields significantly improved coincidence rates of up to plus 33.2\%.

The remainder of this work is structured as follows. 
\autoref{sec:Background} reviews the basic formalism of LOQC.
\autoref{sec:Motivation} motivates the proposed compilation strategy before describing its working principles in \autoref{sec: Proposed Compilation Strategy}. 
An evaluation of the results obtained from simulated experiments is provided by \autoref{sec:Evaluation}, followed by a conclusion of the work in \autoref{sec:Conclusion}.

\section{Background}
\label{sec:Background}

This section lays the foundation for the formalism and terminology used throughout this work. 

\subsection{Formalism for Linear Optics}
\label{sec:LOQC}

We consider photonic experiments in which indistinguishable photons enter a linear optical system, referred to as the linear optical circuit, with $m$ input and $m$ output ports, followed by single-photon detectors. Formally, an input state with $n_1$ photons in port 1, $n_2$ in port 2, up to $n_m$ for port $m$, is described by the Fock state $\ket{\textbf{n}}=\ket{n_1, n_2, \dots, n_m}$. The linear optical transformation is represented by an $m \times m$ unitary matrix $\hat{U} = (u_{ij})$. This unitary is constructed by multiplying the individual unitaries of the individual components of the circuit.

\begin{figure}
        \centering
        \includegraphics[width=0.9\linewidth]{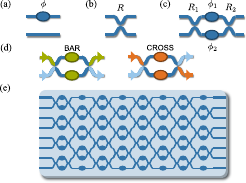}
        \caption{
        \textbf{Building blocks and interferometers for linear optics.} Dark blue lines correspond to waveguides. Light travels from left to right.
        }
        \label{fig:linear_optics}
\end{figure}

\begin{example}
\autoref{fig:linear_optics}a and \autoref{fig:linear_optics}b respectively represent a phase shifter with phase $\phi$ and a beam splitter of reflectivity $R$. Their associated unitary matrices are given by
\begin{equation*}
\hat{U}_\text{PS}(\phi) =
\begin{bmatrix}
    e^{i\phi} & 0 \\
    0 & 1
\end{bmatrix},\quad
\hat{U}_\text{BS}(R) =
\begin{bmatrix}
    \sqrt{R} & i\sqrt{1-R} \\
    i\sqrt{1-R} & \sqrt{R}
\end{bmatrix}.
\end{equation*}
\end{example}

The probability of detecting an output configuration $\ket{\textbf{p}}=\ket{p_1, p_2, \dots, p_m}$ after the linear transformation $\hat{U}$ is given by~\cite{Scheel_2004}
\begin{equation}
\mathbb{P}(\textbf{n} \longrightarrow \textbf{p}) =
\frac{1}{\prod_{j=1}^{m} n_j! \prod_{k=1}^{m} p_k!}
\left|\mathrm{Per}\left[\hat{U}^{\textbf{n}}_{\textbf{p}}\right]\right|^2,
\label{eq:Probability}
\end{equation}
where $\mathrm{Per}$ denotes the matrix permanent, and $\hat{U}^{\textbf{n}}_{\textbf{p}}$ is obtained from $\hat{U}$ by repeating its columns and rows according to the occupation numbers $n_j$ and $p_k$. 

\begin{example}
    By Eq.~\ref{eq:Probability}, a single photon entering input port $i$ is detected in output port $j$ with probability
    \begin{equation}
        \mathbb{P}(\ket{n_i=1} \longrightarrow \ket{n_j=1}) = |u_{ji}|^2.
    \end{equation}
\end{example}

While such experiments can encode qubits, for instance using dual-rail encoding~\cite{Kok2007}, the unitary evolution in the qubit space does not coincide with the matrix $\hat{U}$ acting on the photons.

In practice, the transformation $\hat{U}$ is typically implemented using a linear interferometer composed of fixed beam splitters and reconfigurable phase shifters that are connected via \emph{waveguides}.
Among these implementations, the Mach–Zehnder interferometer serves as a fundamental building block.

\begin{figure*}
    \centering
    \includegraphics[width=\linewidth]{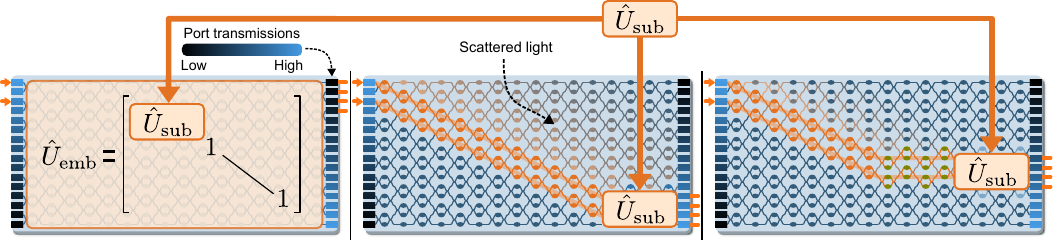}\\
    %\vspace{-16pt}%
    \raggedright%
    \begin{subfigure}{164pt}
    \caption{Embedding}
        \label{fig:basic_compilation}
    \end{subfigure}\hfill%
    \begin{subfigure}{164pt}
    \caption{Port transmission rate maximization}
        \label{fig:input_output_transmissions}
    \end{subfigure}\hfill%
    \begin{subfigure}{164pt}
    \caption{Port transmission rate/routing loss tradeoff}
        \label{fig:input_output_transmissions_routing}
    \end{subfigure}
    \caption{\textbf{Illustration of different methods to implement a 4-by-4 subcircuit on a 16-waveguide Bell-Walmsley circuit.} Orange boxes indicate the placed subcircuit, orange lines the photon routing. Blue bars at the side of the circuits indicate the transmission rates of the respective input and output ports.}
    \vspace{-5mm}
    \label{fig:devices}
\end{figure*}

% \begin{figure*}
%     \centering
%     \includegraphics[width=\linewidth]{figures/subcircuit.pdf}\\
%     \caption{
%     \textbf{Strategies for high-fidelity subcircuit compilation.}
%     }
%     \label{fig:subcircuit}
% \end{figure*}

\subsection{Mach-Zehnder Interferometers}
\label{sec:MZI}

\emph{Mach-Zehnder Interferometers} (MZIs) consist of two beam splitters of reflectivity $R_1$ and $R_2$ enclosing two phase shifters with phase $\phi_1$ and $\phi_2$ as shown in \autoref{fig:linear_optics}c. When the beam splitters are symmetric, that is $R_1=R_2=0.5$, an MZI acts as a beam splitter of tunable reflectivity $R=\sin^2\left(\frac{\phi_1 - \phi_2}{2}\right)$, up to additional phases on input and output ports.

\begin{example}
    Setting  $\phi_1-\phi_2=0$ configures the MZI in the CROSS state, in which light entering one input waveguide exits through the opposite output waveguide. On the contrary, light stays in the same waveguide when $\phi_1-\phi_2=\pm \pi$, which corresponds to the BAR state. The CROSS and BAR configurations are illustrated on~\autoref{fig:linear_optics}d.
\end{example}

In practice, fabrication imperfections lead to beam splitter reflectivity errors, i.e., $R_1,R_2\neq 0.5$. As a result, the ideal CROSS and BAR configurations cannot be exactly realized~\cite{Bandyopadhyay2021}. Nevertheless, the BAR configuration is typically implemented more accurately than the CROSS configuration, as fabrication errors tend to be spatially correlated, leading to similar reflectivities for the two beam splitters within an MZI.

Arrays of MZIs can be programmed to route optical signals between input and output waveguides, as in field-programmable photonic circuits~\cite{Bogaerts2020}, by selecting appropriate sequences of BAR and CROSS states. By interleaving MZIs with additional phase shifters, universal interferometer meshes, such as the Bell-Walmsley architecture~\cite{Bell2021} displayed in~\autoref{fig:linear_optics}.e, can be realized. These meshes can implement an $m \times m$ unitary matrix, or equivalently, any linear optical transformation acting on $m$ waveguides. The operation of such interferometers relies on numerical methods that convert a target unitary matrix into a corresponding set of phase shifts applied to the phase shifters \cite{Reck1994, Clements2016, Bell2021, Kumar2021, Bandyopadhyay2021, Mower2015, Pai2019, Fyrillas2024}.

\section{Motivation}
\label{sec:Motivation}
In contrast to the ever-growing number of compilers for gate-based quantum computing architectures~\cite{yannick, 10993241, 10783033, 11250295, Schmid_2024, 10.1145/3297858.3304075, 10.1145/3297858.3304023, zhu2025quantumcompilerdesignqubit}, LOQC still lacks efficient compilers for various compilation tasks.
In this work, we consider the task of implementing a \emph{target unitary matrix} that represents the desired computation on the circuit.
To this end, the phase shifters on the circuit must be tuned such that the resulting unitary matrix describing the circuit's behavior is as close as possible to the target unitary matrix.
Note that in the frame of this work, we consider the Bell-Walmsley architecture~\cite{Bell2021} for the linear optical circuit.

Existing compilation solutions assume that the target unitary spans the same dimensions as the circuit.
For a smaller target unitary $\hat{U}_\text{sub}$, referred to as the \emph{subcircuit}, this assumption can be satisfied by expanding it with ones on the diagonal and zeros elsewhere until the circuit size is reached, yielding an embedded target unitary $\hat{U}_\text{emb}$.
The compiler then tries to find parameters $\vec{\phi}$ for the phase shifters that maximize the overlap between $\hat{U}_\text{emb}$ and the unitary implemented by the circuit.

To this end, a unitary parameterized by the phase shifters $\hat{U}(\vec{\phi})$ that describe the circuit's behavior can be generated and, via for instance gradient descent, $\vec\phi$ can be determined so that $\hat{U}(\vec{\phi})$ is as close as possible to $\hat{U}_\text{emb}$. 
This solution may use large parts of the circuit to implement the desired computation, effectively diluting the smaller initial target unitary $\hat{U}_\text{sub}$ across the entire circuit.

\begin{example}
\label{ex:baseline}
    The illustration in \autoref{fig:basic_compilation} shows an implementation of a 4-by-4 target unitary subcircuit $\hat{U}_\text{sub}$ on a Bell-Walmsley circuit with 16 waveguides. Using the existing compilation workflows mentioned above, the computation zone (orange) is spread across the entire circuit.
\end{example}

However, given the reduced size of the subcircuit, the computation can be realized on a fraction of the entire circuit.
We call this fraction the \emph{computation zone}.
This computation zone can be arbitrarily placed on the circuit, like a small box in a larger one. 
% \footnote{Note that in the frame of this work, it is considered to always place the computation zone at the end of the circuit, as this does not restrict the quality of the results.}
Hence, the compiler is given optimization opportunities that currently available approaches miss, which, if appropriately exploited, can tackle major engineering challenges in LOQC.
That is, circuit imperfections must be handled wisely during the compilation to fully exploit the hardware's capabilities.

More precisely, one major challenge in LOQC with respect to hardware imperfections is the transmission rate, i.e., the ratio of light exiting the circuit to that entering the circuit.
Given that many LOQC computations are inherently probabilistic, the transmission rate is a crucial factor for maintaining efficiency at scale.
This can be seen in the coincidence rate (the probability of measuring the desired $n$ photons at the same time), which scales proportionally to $T^n$ with the transmission rate $T$ and the number of photons $n$.
Together with the probabilistic nature of many LOQC experiments \cite{Ralph2002, Knill2002}, this exponential behavior underlines the importance of increasing the transmission rate for successful scaling.

Consider that, due to hardware imperfections, each input and output port has a different transmission rate.
Hence, by placing the computation zone wisely and thus using a combination of high-quality input and output ports, the overall coincidence rate of the resulting computation can be improved.

\begin{example}
     \autoref{fig:input_output_transmissions} illustrates a solution that keeps the computation zone (orange box) compact instead of using the entire circuit. Photons are actively routed (orange lines) to this computation zone. This way, input and output ports that are not in line but have good transmission rates are connected via the routing part, while still leaving enough resources for the actual computation. However, this simplistic approach overlooks that photon fractions are lost due to routing operations, as indicated by the orange gradient in the figure and explained in the following.
\end{example}

Despite this more sophisticated approach, further effects on the circuit should be considered to improve the result.
Due to imperfect beam splitters, routing operations can also degrade transmission rates, with CROSS (switching waveguides) movements showing stronger degradation than BAR (staying on the same waveguide) in current hardware.
Even though these photons are technically not actually lost, they can fail to enter the computation zone as desired, which needs to be accounted for.
Hence, the resulting goal is to choose a high-quality set of input and output ports while keeping routing losses low.

\begin{example}
    The last example in \autoref{fig:input_output_transmissions_routing} shows another possible solution that also accounts for routing losses. Since CROSS movements can cause more photons to leak out of the desired routing lane than BAR movements, it may be better overall to choose slightly worse output ports to save on routing costs.
\end{example}

Overall, the compilation step has significant potential that needs to be unlocked to fully overcome engineering challenges in LOQC.
However, considering and balancing the various physical aspects of the circuit to find a good solution is highly non-trivial.
To automate this process, this work proposes a method that accounts for input and output transmission rates as well as the individual beam splitter reflectivities to determine a good solution with the click of a button.

\section{Proposed Compilation Strategy}
\label{sec: Proposed Compilation Strategy}

This section describes the approach for an automated compiler for placing and routing a subcircuit on a linear optical circuit, accounting for input and output port transmission rates and beam splitter reflectivities to optimize for coincidence rate.
Given the mesh-like architecture of the linear optical circuit, the problem is mapped to a graph.
Photon routing and subcircuit placement are then extracted from the graph.
To this end, edge weights are chosen corresponding to transmission costs, and the route with the lowest cost is determined.
%Subsequently, this route is translated into a phase-matrix, representing the phase shifters \yannick[in]{Should it be ``on''?} the circuit.
Subsequently, a phase-matrix is created, representing the transformation implemented by the circuit's phase shifters, with constraints corresponding to the chosen route.
In a step called \emph{phase compilation}, the phases of this matrix are adjusted, in particular the ones of the computation zone, so that the circuit's behavior matches the target unitary as closely as possible while preserving the routing constraints.
The necessary steps are described in more detail below.

\subsection{Modeling the Problem}

%To enable efficient mapping to a graph, an abstract model of the problem is required that aims to keep the resulting problem complexity low.
In the considered scenario, the photons enter the linear optical circuit into even input ports as in~\cite{Kok2007}.
Hence, the input state of the photons (for the considered input ports) always equals the Fock state $\ket{1,0,1,0,...,0}$ of length equal to the target unitary's dimension.
These photons are kept adjacent to each other during routing before they enter the computation zone, which is always placed at the end of the circuit.
Depending on the routing, the photons may enter the computation zone as $\ket{1,0,1,0,...,0}$ or $\ket{0,1,0,1,...,1}$.
In the latter case, given the dependency of the result on the input state, this must be corrected by multiplying the target unitary by a permutation matrix, effectively swapping every odd column with its adjacent even column.
%This product is taken as an effective target unitary for phase compilation to overall reflect the originally desired semantics.

\subsection{Graph}

The structure of the linear optical circuit is directly reflected in the graph modeling the problem.
As described in \autoref{sec:Background}, the circuit consists of a mesh of waveguides arranged in rows, with successive layers of MZIs that couple exactly two adjacent waveguides.
The graph mirrors this structure: each layer of nodes corresponds to a layer of MZIs, and each node represents an individual waveguide in that layer.
Edges connect nodes of adjacent layers and encode the two possible switching states of each MZI (BAR and CROSS), with edge weights corresponding to the losses due to routed photons.
Since signal propagation through the circuit is strictly forward, all edges point from one layer to the next, making the graph a \emph{directed acyclic graph}~(DAG).

\begin{figure}
    \centering
    \includegraphics[width=0.95\linewidth]{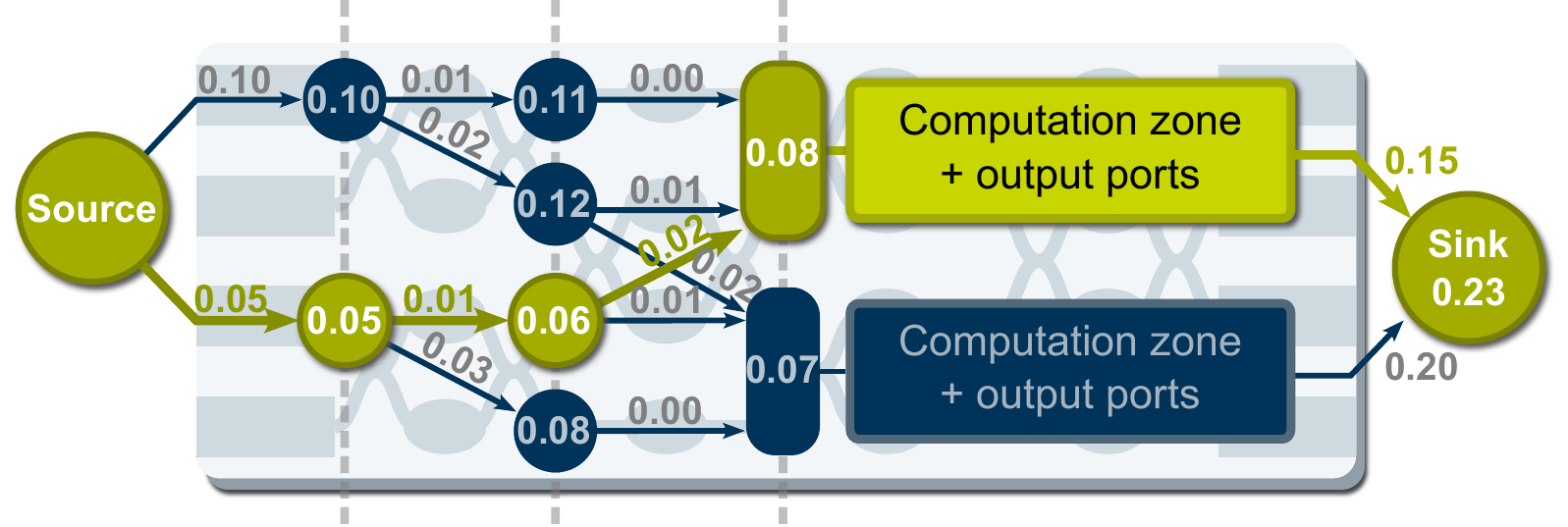}
    \caption{
        \textbf{Problem Graph.}
       The depicted graph corresponds to a 4-waveguide Bell-Walmsley circuit and a 2-by-2 target unitary.
    }
    \label{fig:graph}
\end{figure}

\begin{example}
    \autoref{fig:graph} shows such a graph for a small example of a 4-waveguide Bell-Walmsley circuit and a 2-by-2 target unitary (hence one photon).
    The dashed lines indicate the layers, the circles their nodes, and the arrows their edges.
\end{example}

Since all injected photons are routed in parallel, the possible routes through the graph correspond to the route of the uppermost photon.
The other photons follow the turn of the first one and are not directly represented in the graph.
Similar to well-established graph algorithms, the one employed in this work adds the costs of edges for a specific route.
Since transmissions are multiplied and in the interval \([0,1]\), the weights are expressed on a negative logarithmic scale.
This way, their sum corresponds to multiplying the transmission, and the algorithm can minimize the additive cost as usual.
The graph's edges from and to the source and sink nodes correspond to the in- and output ports.

\begin{example}
    \autoref{fig:graph} shows the edge weights on top of each edge.
    The source node and its edges correspond to the negative logarithmic input port transmission costs.
    The computation zones are part of the last edges ending in the sink node.
    Their cost combines the costs of all the computation zone's output ports (2 in the example). %, since photons can exit the computation zone through each of these ports.
\end{example}

\subsection{Determine Cheapest Route}

Standard shortest-route algorithms such as Dijkstra's or A* are general-purpose methods that do not assume any particular graph structure.
Applying them here would therefore ignore two key properties of the graph: its strict layering and the absence of cycles.
Exploiting these properties enables a more efficient, exact solution based on dynamic programming outlined in \autoref{alg:shortest_path}.

The algorithm initializes the source node with a cost of zero (line 1) and all other nodes with $\infty$ (line 2-3).
The layers are then processed in topological order (line 4).
For each node $u$ in the current layer (line 5), the algorithm examines all outgoing edges $(u, v)$ (line 6) and computes the accumulated cost $c = d(u) + w(u, v)$ (line 7).
If $c$ is smaller than $v$'s current tentative cost (line 8), both the cost and the predecessor pointer are updated (line 9).
Because the graph is a DAG with fixed layering, each node is visited exactly once, and its optimal cost is finalized before any of its successors are processed.
After termination, every node holds the cost of the cheapest route from the source, and the \mbox{source-to-sink} route can be recovered by backtracking through the predecessor pointers.

\begin{algorithm}[t]
\caption{Cheapest Route in a Layered DAG}\label{alg:shortest_path}
\KwIn{Layered DAG $G = (V, E, w)$, ordered layer sequence
      $\mathcal{L} = (L_1, L_2, \dots, L_K)$, source node $s \in L_1$}
\KwOut{Distance map $d: V \to \mathbb{R}_{\geq 0} \cup \{\infty\}$,
       predecessor map $\pi: V \to V \cup \{\text{NIL}\}$}
$d(s) \gets 0$; $\pi(s) \gets \text{NIL}$\;
\ForEach{$v \in V \setminus \{s\}$}{
  $d(v) \gets \infty$; $\pi(v) \gets \text{NIL}$\;
}
\ForEach{$L_k \in \mathcal{L}$}{
  \ForEach{$u \in L_k$}{
    \ForEach{$(u, v) \in E$}{
      $c \gets d(u) + w(u, v)$\;
      \If{$c < d(v)$}{
        $d(v) \gets c$; $\pi(v) \gets u$\;
      }
    }
  }
}
\end{algorithm}

\begin{example}
    The green route in \autoref{fig:graph} highlights the cheapest route through the graph.
    After assigning costs to reach each node using the \autoref{alg:shortest_path}, the cheapest route is determined by following the predecessor pointers from the sink.
\end{example}

Based on this route, a phase matrix is determined that imposes constraints on certain phase shifters to satisfy the routing conditions.
Note that the routing operations only dictate the internal phase difference of the two phase shifters within an MZI, not their absolute values, as described in \autoref{sec:Background}.
% Additionally, phase shifters not required for routing or computation can be kept inactive.

\begin{table*}[t]
    \centering
    \small
    \setlength{\tabcolsep}{4.8pt} % Narrower column spacing
    \caption{Results: colored values indicate the difference between the proposed approach and the baseline (\textcolor{my_green}{green} = improvement, \textcolor{my_red}{red} = degradation). For each scenario, the results were averaged over the five Haar-random target unitaries.}
    \label{tab:results}
    \begin{tabular}{
        S[table-format=1.3]% Assumed Phase Errors
        S[table-format=2.0]% Circuit Dimension
        S[table-format=1.0]% Subcircuit Dimension
        S[table-format=2.2]% Baseline Compilation Time
        S[exponent-mode=scientific, table-format=1.1e2, round-mode=figures, round-precision=2]% Baseline TVD
        S[table-format=1.3]% Baseline Transmission
        S[table-format=2.2]% Proposed Compilation Time
        S[exponent-mode=scientific, table-format=1.1e2, round-mode=figures, round-precision=2]% Proposed TVD
        @{\ (\hspace{.2ex}}r@{)\hspace{2ex}} % Delta
        S[table-format=1.3]% Proposed Transmission
        @{\ (\hspace{.2ex}}r@{)\hspace{2ex}} % Delta
        S[table-format=1.3]% Proposed Transmission
    }
        \toprule
        {Phase Noise} & {Circuit} & {Subcircuit} & \multicolumn{3}{c}{\textbf{Baseline}} & \multicolumn{5}{c}{\textbf{Proposed Method}} \\[-0.6ex]
        \cmidrule(lr){4-6} \cmidrule(l){7-11}
        {[mrad]} & {Dim.} & {Dim.} & {Comp. [\si{\minute}]} & {TVD} & {Coinc. Rate} & {Comp. [\si{\minute}]} & \multicolumn{2}{c}{TVD} & \multicolumn{2}{c}{Coinc. Rate} \\
        \midrule
        0 & 24 & 4 & 0.38 & 0.000007884 & 0.562 & 0.61 & 0.002739795 & \ptvd{ 0.003} & 0.844 & \ptrans{0.282} \\
              &    & 8 & 0.64 & 0.000119177 & 0.265 & 0.78 & 0.034403161 & \ptvd{ 0.034} & 0.425 & \ptrans{0.161} \\[0.6ex]
              & 48 & 4 & 0.80 & 0.000023920 & 0.471 & 2.85 & 0.001571113 & \ptvd{ 0.002} & 0.799 & \ptrans{0.327} \\
              &    & 8 & 1.85 & 0.000044893 & 0.335 & 3.41 & 0.012129443 & \ptvd{ 0.012} & 0.481 & \ptrans{0.146} \\[1.2ex]
        15 & 24 & 4 & 0.38 & 0.026820520 & 0.560 & 0.61 & 0.023399629 & \ptvd{-0.003} & 0.841 & \ptrans{0.280} \\
              &    & 8 & 0.64 & 0.058717258 & 0.264 & 0.78 & 0.058529423 & \ptvd{ 0.000} & 0.424 & \ptrans{0.161} \\[0.6ex]
              & 48 & 4 & 0.77 & 0.036043667 & 0.469 & 2.80 & 0.027130628 & \ptvd{-0.009} & 0.796 & \ptrans{0.327} \\
              &    & 8 & 1.80 & 0.090833230 & 0.332 & 3.30 & 0.049926471 & \ptvd{-0.041} & 0.472 & \ptrans{0.140} \\[1.2ex]
        30 & 24 & 4 & 0.38 & 0.052009342 & 0.558 & 0.61 & 0.035113407 & \ptvd{-0.017} & 0.832 & \ptrans{0.274} \\
              &    & 8 & 0.69 & 0.122503013 & 0.260 & 0.83 & 0.109203353 & \ptvd{-0.013} & 0.414 & \ptrans{0.153} \\[0.6ex]
              & 48 & 4 & 0.79 & 0.084370573 & 0.454 & 2.83 & 0.080987676 & \ptvd{-0.003} & 0.786 & \ptrans{0.332} \\
              &    & 8 & 1.80 & 0.167065881 & 0.325 & 3.30 & 0.090683148 & \ptvd{-0.076} & 0.457 & \ptrans{0.132} \\
        \bottomrule
    \end{tabular}
    \vspace{-2mm}
\end{table*}

\subsection{Phase Compilation}

To conduct the desired computation on the circuit, the phase shifters must be tuned in a way such that the unitary matrix representing the circuit's behavior is as close as possible to the target unitary $\hat{U}_\text{sub}$.
By multiplying the unitary matrices of the individual beam splitters and phase shifters of the circuit, a parameterized unitary of the circuit's behavior $\hat{U}(\vec{\phi})$ is derived.
These parameters $\vec{\phi}$ are adjusted via gradient descent until the overlap of the two unitaries ($\hat{U}(\vec{\phi})$ and $\hat{U}_\text{sub}$) is as large as possible~\cite{phase_compilation}.
To implement the desired route determined by the graph algorithm, certain internal phase shifter differences within an MZI must be satisfied, yet their sum is a degree of freedom for the optimizer.
This can be used to partially compensate for the negative effects on photons due to routing.

\section{Evaluation}
\label{sec:Evaluation}

The proposed methods were implemented and are publicly available as open-source software as part of the Munich Quantum Toolkit~\cite{mqt} within MQT QMAP\footnote{\url{https://github.com/munich-quantum-toolkit/qmap}}.
To demonstrate the compiler's effectiveness, a comprehensive evaluation was conducted by compiling subcircuits to linear optical circuits, each of varying sizes.
The results were evaluated with simulations of each scenario.
Before discussing the results, the experimental setup is briefly described.

\subsection{Setup}

The evaluations consider linear optical circuits based on the Bell-Walmsley architecture, consisting 
of either 24 or 48 waveguides.
On those, subcircuit computations implementing target unitaries with dimensions 4 and 8, respectively, are realized.
To this end, for each scenario, five Haar-random target unitaries were generated, and for each, placement and routing were determined and phases compiled using the proposed method.
For comparison, the target unitaries were additionally compiled with a baseline method that embeds the target unitary into a circuit-sized unitary as described in \autoref{sec:Motivation}.
This effectively selects the first 4 and 8 input and output ports, respectively, entirely agnostic of their actual transmission rates.

To model realistic device behavior, transmission rates of in- and output ports were drawn from a uniform distribution between 0.7 and 1.0 before normalization, representing realistic values.
Beam splitter reflectivities are taken from an actual Quandela device for the 24-waveguide circuit, and random values of comparable magnitude for the 48-waveguide circuit.
For each benchmark, we compare three different scenarios, with phase noise drawn from a normal distribution with standard deviations of 0, 15, and 30 mrad.
The approach easily generalizes to any parameter combination, allowing the generated parameters to be replaced with real calibration data.

To evaluate the compilation results, the circuits were simulated for each combination of target 
unitary, circuit size, and phase noise level using Perceval~\cite{Heurtel2023percevalsoftware} 
(v1.1.0) in Python (v3.13.11) on an Apple MacBook Pro (M3 Pro, 36 GB RAM).
We assess each scenario using three figures of merit: compilation time, fidelity in the form of the \emph{total variation distance}~(TVD)---a standard distance measure for discrete probability distributions~\cite{6804281}---between simulated and ground truth output distributions given the input state $\ket{1,0,1,0,...,0}$, and the coincidence rate, defined as the probability of measuring the desired number of photons in the considered output ports.
The results are summarized in \autoref{tab:results}. %, structured by phase error, circuit dimension, and subcircuit dimension.

\subsection{Compilation Time}

In general, the compilation time of the proposed approach is larger in all cases but remains in the same order of magnitude.
This is expected, as the compilation involves more steps that explicitly exploit the optimization opportunities discussed in \autoref{sec:Motivation} and eventually provides results of higher quality (as also confirmed by the results in the next sections). 
More precisely, while the baseline merely compiles the phases for the given circuit, the proposed approach also generates the graph, finds the best route, translates it into predefined phases, and subsequently performs phase compilation. 
However, using a graph algorithm tailored to exploit the properties of the directed acyclic graph renders the additional time costs negligible.
Since the phase compilation also considers the entire circuit (to partially correct for routing-induced effects on the photons), its runtime is comparable to the baseline.

\subsection{Total Variation Distance (TVD)}

When comparing the TVD, the proposed approach shows weaker performance without phase noise and better performance with two levels of phase noise.
In the unrealistic case of no phase noise, the baseline does not face a noticeable source of error, resulting in accurate computations. 
Given the routing-induced effects on the photons in the proposed approach, such as leaked photons from one path into another, it cannot reach the same accuracy without phase noise.
However, under realistic phase noise, the proposed approach is more robust, as it uses fewer phase shifters for computation, thereby reducing error accumulation.
This can be observed in \autoref{tab:results} with the increasing difference in TVD between the two approaches with increasing phase noise.
This robustness cannot be achieved with the baseline, which effectively spreads the computation zone across a large area of the circuit.
As a result, it uses more phase shifters, which accumulate more phase errors, leading to lower accuracy.
Crucially, robustness to phase noise is an essential ingredient for scaling LOQC, as errors accumulate with increasing circuit size, thereby limiting scalability with existing compilation methods.
The proposed approach directly addresses this by keeping the computation zone compact, thereby bounding error accumulation and enabling scaling.

\subsection{Coincidence Rate}

Overall, under realistic phase noise, beam splitter reflectivities, and port transmission parameters, the proposed method consistently outperforms the baseline in coincidence rates across all considered benchmarks.
As indicated by the consistently green values in \autoref{tab:results}, the method yields substantially higher values of up to plus 33.2\%.
Unlike existing methods, the proposed approach explicitly accounts for the input and output port transmissions and beam splitter reflectivities during compilation that critically influence the resulting coincidence rate.
This matters substantially in linear optical quantum computing, since coincidence rates directly govern the rate at which valid multi-photon events are detected.
Given that LOQC gates are inherently probabilistic, low coincidence rates translate to exponentially longer experiment times and fundamentally limit the scalability of quantum circuits.
The significant improvement in coincidence rate achieved by the proposed method thus leads to substantially faster computations and, crucially, enables the scaling of LOQC experiments that would otherwise be rendered impractical by photon loss.

\section{Conclusion}
\label{sec:Conclusion}

Recent advances in hardware quality and scaling have significantly improved \emph{Linear Optical Quantum Computing} (LOQC), although important challenges remain.
One major challenge is the circuit's transmission rate, i.e., the ratio of light exiting the circuit to that injected, which is affected by imperfect components such as photon-loss-affected input and output ports.
When performing a computation smaller than the circuit, a subcircuit, one can choose where to conduct it on the circuit, thereby combining high-quality input and output ports.
However, photons must be actively routed to this zone, which introduces negative effects due to imperfect beam splitters in the circuit, which must be accounted for.
This work proposes the first automated compiler for subcircuits in LOQC, tailored to optimize coincidence rates---the probability of measuring the desired number of photons---by accounting for imperfect input and output ports and beam splitter reflectivities.
Compared to existing methods, the proposed approach shows greater robustness against phase noise and significantly improved coincidence rates in simulated experiments.
Future work includes demonstrating the approach's effectiveness on a real LOQC device and extending it to include additional circuit imperfections.

\section*{Acknowledgments}

\small
The project leading to this publication has received funding from the European Research Council (ERC) under the European Union’s Horizon 2020 research and innovation program (grant agreement No. 101001318). Furthermore, this work is part of the Munich Quantum Valley, which is supported by the Bavarian state government with funds from the Hightech Agenda Bayern Plus.

Additionally, this work has been funded by the TUF-TOPIQC project part of the Trilateral Call for Quantum Innovation, co-financed by Germany, the Netherlands and by the French National Quantum Strategy (France 2030) program.

During the preparation of the code and manuscript, the authors used GitHub Copilot, powered by OpenAI's GPT Codex 5.3 and Google's Gemini 3 to improve code, spelling, grammar, clarity, and readability. Afterward, the authors reviewed and edited the content as needed. The authors take full responsibility for the final content.

\vspace{2000pt}

\renewcommand*{\bibfont}{\small}
\setlength{\bibitemsep}{0.3em}
\printbibliography

@STRING{tcad	= {{IEEE} Trans. on {CAD} of Integrated Circuits and Systems} }

@STRING{is	= {IEEE Software} }

@STRING{iccad	= {Int'l Conf. on CAD} }

@STRING{date	= {Design, Automation and Test in Europe} }

@STRING{asplos = {{Int'l Conf. on Architectural Support for Programming Languages and Operating Systems}} }

@STRING{qsw = {{Int'l Conf. on Quantum Software}} }

@STRING{qce = {{Int'l Conf. on Quantum Computing and Engineering}} }

@article{Aaronson2013, title={The Computational Complexity of Linear Optics}, volume={9}, DOI={10.4086/toc.2013.v009a004}, number={4}, journal={Theory of Computing}, publisher={Theory of Computing}, author={Aaronson, Scott and Arkhipov, Alex}, year={2013}, month=feb, pages={143–252} }

@article{Aspuru-Guzik_Walther_2012, title={Photonic quantum simulators}, volume={8}, rights={2012 Springer Nature Limited}, ISSN={1745-2481}, DOI={10.1038/nphys2253}, number={4}, journal={Nature Physics}, publisher={Nature Publishing Group}, author={Aspuru-Guzik, Alán and Walther, Philip}, year={2012}, month=apr, pages={285–291},  }

@article{Bandyopadhyay2021, title={Hardware error correction for programmable photonics}, volume={8}, rights={&#169; 2021 Optical Society of America}, ISSN={2334-2536}, DOI={10.1364/OPTICA.424052}, number={10}, journal={Optica}, publisher={Optica Publishing Group}, author={Bandyopadhyay, Saumil and Hamerly, Ryan and Hamerly, Ryan and Englund, Dirk}, year={2021}, month=oct, pages={1247–1255},  }

@article{Bell2021, title={Further compactifying linear optical unitaries}, volume={6}, ISSN={2378-0967}, DOI={10.1063/5.0053421}, number={7}, journal={APL Photonics}, author={Bell, B. A. and Walmsley, I. A.}, year={2021}, month=july, pages={070804} }

@article{Bogaerts2020, title={Programmable photonic circuits}, volume={586}, rights={2020 Springer Nature Limited}, ISSN={1476-4687}, DOI={10.1038/s41586-020-2764-0}, number={78287828}, journal={Nature}, publisher={Nature Publishing Group}, author={Bogaerts, Wim and Pérez, Daniel and Capmany, José and Miller, David A. B. and Poon, Joyce and Englund, Dirk and Morichetti, Francesco and Melloni, Andrea}, year={2020}, month=oct, pages={207–216},  }

@article{Brod2019, title={Photonic implementation of boson sampling: a review}, volume={1}, ISSN={2577-5421, 2577-5421}, DOI={10.1117/1.AP.1.3.034001}, number={3}, journal={Advanced Photonics}, publisher={SPIE}, author={Brod, Daniel J. and Galvão, Ernesto F. and Crespi, Andrea and Osellame, Roberto and Spagnolo, Nicolò and Sciarrino, Fabio}, year={2019}, month=may, pages={034001} }

@article{Cerezo2021, title={Variational quantum algorithms}, volume={3}, rights={2021 Springer Nature Limited}, ISSN={2522-5820}, DOI={10.1038/s42254-021-00348-9}, number={9}, journal={Nature Reviews Physics}, publisher={Nature Publishing Group}, author={Cerezo, M. and Arrasmith, Andrew and Babbush, Ryan and Benjamin, Simon C. and Endo, Suguru and Fujii, Keisuke and McClean, Jarrod R. and Mitarai, Kosuke and Yuan, Xiao and Cincio, Lukasz and Coles, Patrick J.}, year={2021}, month=sept, pages={625–644},  }

@article{Chi2022, title={A programmable qudit-based quantum processor}, volume={13}, rights={2022 The Author(s)}, ISSN={2041-1723}, DOI={10.1038/s41467-022-28767-x}, number={1}, journal={Nature Communications}, publisher={Nature Publishing Group}, author={Chi, Yulin and Huang, Jieshan and Zhang, Zhanchuan and Mao, Jun and Zhou, Zinan and Chen, Xiaojiong and Zhai, Chonghao and Bao, Jueming and Dai, Tianxiang and Yuan, Huihong and Zhang, Ming and Dai, Daoxin and Tang, Bo and Yang, Yan and Li, Zhihua and Ding, Yunhong and Oxenløwe, Leif K. and Thompson, Mark G. and O’Brien, Jeremy L. and Li, Yan and Gong, Qihuang and Wang, Jianwei}, year={2022}, month=mar, pages={1166},  }

@article{Clements2016, title={Optimal design for universal multiport interferometers}, volume={3}, ISSN={2334-2536}, DOI={10.1364/OPTICA.3.001460}, number={12}, journal={Optica}, publisher={Optica Publishing Group}, author={Clements, William R. and Humphreys, Peter C. and Metcalf, Benjamin J. and Kolthammer, W. Steven and Walmsley, Ian A.}, year={2016}, month=dec, pages={1460–1465},  }

@article{Deng2023, title={Gaussian Boson Sampling with Pseudo-Photon-Number-Resolving Detectors and Quantum Computational Advantage}, volume={131}, DOI={10.1103/PhysRevLett.131.150601}, number={15}, journal={Physical Review Letters}, publisher={American Physical Society}, author={Deng, Yu-Hao and Gu, Yi-Chao and Liu, Hua-Liang and Gong, Si-Qiu and Su, Hao and Zhang, Zhi-Jiong and Tang, Hao-Yang and Jia, Meng-Hao and Xu, Jia-Min and Chen, Ming-Cheng and Qin, Jian and Peng, Li-Chao and Yan, Jiarong and Hu, Yi and Huang, Jia and Li, Hao and Li, Yuxuan and Chen, Yaojian and Jiang, Xiao and Gan, Lin and Yang, Guangwen and You, Lixing and Li, Li and Zhong, Han-Sen and Wang, Hui and Liu, Nai-Le and Renema, Jelmer J. and Lu, Chao-Yang and Pan, Jian-Wei}, year={2023}, month=oct, pages={150601} }

@article{Dong2017, title={Silicon-on-Insulator Waveguide Devices for Broadband Mid-Infrared Photonics}, volume={9}, ISSN={1943-0655}, DOI={10.1109/JPHOT.2017.2692039}, number={3}, journal={IEEE Photonics Journal}, author={Dong, Bowei and Guo, Xin and Ho, Chong Pei and Li, Bo and Wang, Hong and Lee, Chengkuo and Luo, Xianshu and Lo, Guo-Qiang}, year={2017}, month=june, pages={1–10} }

@article{Flamini2018, title={Photonic quantum information processing: a review}, volume={82}, ISSN={0034-4885}, DOI={10.1088/1361-6633/aad5b2}, number={1}, journal={Reports on Progress in Physics}, publisher={IOP Publishing}, author={Flamini, Fulvio and Spagnolo, Nicolò and Sciarrino, Fabio}, year={2018}, month=nov, pages={016001},  }

@article{Fyrillas2024, title={Scalable machine learning-assisted clear-box characterization for optimally controlled photonic circuits}, volume={11}, rights={© 2024 Optica Publishing Group}, ISSN={2334-2536}, DOI={10.1364/OPTICA.512148}, number={3}, journal={Optica}, publisher={Optica Publishing Group}, author={Fyrillas, Andreas and Faure, Olivier and Maring, Nicolas and Senellart, Jean and Belabas, Nadia}, year={2024}, month=mar, pages={427–436},  }

@article{Harris2014, title={Efficient, compact and low loss thermo-optic phase shifter in silicon}, volume={22}, rights={© 2014 Optical Society of America}, ISSN={1094-4087}, DOI={10.1364/OE.22.010487}, number={9}, journal={Optics Express}, publisher={Optica Publishing Group}, author={Harris, Nicholas C. and Ma, Yangjin and Mower, Jacob and Baehr-Jones, Tom and Englund, Dirk and Hochberg, Michael and Galland, Christophe}, year={2014}, month=may, pages={10487–10493},  }

@article{Harris2018, title={Linear programmable nanophotonic processors}, volume={5}, rights={&#169; 2018 Optical Society of America}, ISSN={2334-2536}, DOI={10.1364/OPTICA.5.001623}, number={12}, journal={Optica}, publisher={Optica Publishing Group}, author={Harris, Nicholas C. and Carolan, Jacques and Bunandar, Darius and Prabhu, Mihika and Hochberg, Michael and Baehr-Jones, Tom and Fanto, Michael L. and Smith, A. Matthew and Tison, Christopher C. and Alsing, Paul M. and Englund, Dirk}, year={2018}, month=dec, pages={1623–1631},  }

@article{Kok2007, title={Linear optical quantum computing with photonic qubits}, volume={79}, DOI={10.1103/RevModPhys.79.135}, number={1}, journal={Reviews of Modern Physics}, publisher={American Physical Society}, author={Kok, Pieter and Munro, W. J. and Nemoto, Kae and Ralph, T. C. and Dowling, Jonathan P. and Milburn, G. J.}, year={2007}, month=jan, pages={135–174} }

@article{Knill2002, title={Quantum gates using linear optics and postselection}, volume={66}, DOI={10.1103/PhysRevA.66.052306}, number={5}, journal={Physical Review A}, publisher={American Physical Society}, author={Knill, E.}, year={2002}, month=nov, pages={052306} }

@misc{Kumar2021,
      title={Mitigating linear optics imperfections via port allocation and compilation}, 
      author={Shreya P. Kumar and Leonhard Neuhaus and Lukas G. Helt and Haoyu Qi and Blair Morrison and Dylan H. Mahler and Ish Dhand},
      year={2021},
      eprint={2103.03183},
      archivePrefix={arXiv},
      url={https://arxiv.org/abs/2103.03183}, 
}

@article{Maring2024, title={A versatile single-photon-based quantum computing platform}, rights={2024 The Author(s)}, ISSN={1749-4893}, DOI={10.1038/s41566-024-01403-4}, journal={Nature Photonics}, publisher={Nature Publishing Group}, author={Maring, Nicolas and Fyrillas, Andreas and Pont, Mathias and Ivanov, Edouard and Stepanov, Petr and Margaria, Nico and Hease, William and Pishchagin, Anton and Lemaître, Aristide and Sagnes, Isabelle and Au, Thi Huong and Boissier, Sébastien and Bertasi, Eric and Baert, Aurélien and Valdivia, Mario and Billard, Marie and Acar, Ozan and Brieussel, Alexandre and Mezher, Rawad and Wein, Stephen C. and Salavrakos, Alexia and Sinnott, Patrick and Fioretto, Dario A. and Emeriau, Pierre-Emmanuel and Belabas, Nadia and Mansfield, Shane and Senellart, Pascale and Senellart, Jean and Somaschi, Niccolo}, year={2024}, month=mar, pages={1–7},  }

@article{Mezher2023, title={Solving graph problems with single photons and linear optics}, volume={108}, DOI={10.1103/PhysRevA.108.032405}, number={3}, journal={Physical Review A}, publisher={American Physical Society}, author={Mezher, Rawad and Carvalho, Ana Filipa and Mansfield, Shane}, year={2023}, month=sept, pages={032405} }

@article{Mower2015, title={High-fidelity quantum state evolution in imperfect photonic integrated circuits}, volume={92}, DOI={10.1103/PhysRevA.92.032322}, number={3}, journal={Physical Review A}, publisher={American Physical Society}, author={Mower, Jacob and Harris, Nicholas C. and Steinbrecher, Gregory R. and Lahini, Yoav and Englund, Dirk}, year={2015}, month=sept, pages={032322} }

@article{Pai2019,
  title = {Matrix Optimization on Universal Unitary Photonic Devices},
  author = {Pai, Sunil and Bartlett, Ben and Solgaard, Olav and Miller, David A. B.},
  journal = {Phys. Rev. Appl.},
  volume = {11},
  pages = {064044},
  numpages = {18},
  year = {2019},
  month = {Jun},
  publisher = {American Physical Society},
  doi = {10.1103/PhysRevApplied.11.064044},
  url = {https://link.aps.org/doi/10.1103/PhysRevApplied.11.064044}
}

@article{Qiang2018, title={Large-scale silicon quantum photonics implementing arbitrary two-qubit processing}, volume={12}, rights={2018 The Author(s)}, ISSN={1749-4893}, DOI={10.1038/s41566-018-0236-y}, number={99}, journal={Nature Photonics}, publisher={Nature Publishing Group}, author={Qiang, Xiaogang and Zhou, Xiaoqi and Wang, Jianwei and Wilkes, Callum M. and Loke, Thomas and O’Gara, Sean and Kling, Laurent and Marshall, Graham D. and Santagati, Raffaele and Ralph, Timothy C. and Wang, Jingbo B. and O’Brien, Jeremy L. and Thompson, Mark G. and Matthews, Jonathan C. F.}, year={2018}, month=sept, pages={534–539},  }

@article{Ralph2002, title={Linear optical controlled-NOT gate in the coincidence basis}, volume={65}, DOI={10.1103/PhysRevA.65.062324}, number={6}, journal={Physical Review A}, publisher={American Physical Society}, author={Ralph, T. C. and Langford, N. K. and Bell, T. B. and White, A. G.}, year={2002}, month=june, pages={062324} }

@article{Reck1994, title={Experimental realization of any discrete unitary operator}, volume={73}, DOI={10.1103/PhysRevLett.73.58}, number={1}, journal={Physical Review Letters}, publisher={American Physical Society}, author={Reck, Michael and Zeilinger, Anton and Bernstein, Herbert J. and Bertani, Philip}, year={1994}, month=july, pages={58–61} }

@misc{Scheel_2004,
      title={Permanents in linear optical networks}, 
      author={Stefan Scheel},
      year={2004},
      eprint={quant-ph/0406127},
      archivePrefix={arXiv},
      url={https://arxiv.org/abs/quant-ph/0406127}, 
}

@article{Skryabin2025, title={Heralded generation of programmable two-qubit entangled states on a linear-optical platform}, volume={3}, rights={© 2025 Optica Publishing Group}, ISSN={2837-6714}, DOI={10.1364/OPTICAQ.546244}, number={2}, journal={Optica Quantum}, publisher={Optica Publishing Group}, author={Skryabin, N. N. and Biriukov, Yu A. and Dryazgov, M. A. and Fldzhyan, S. A. and Zhuravitskii, S. A. and Argenchiev, A. S. and Kondratyev, I. V. and Tsoma, L. A. and Okhlopkov, K. I. and Gruzinov, I. M. and Arsenyev, A. Ya and Taratorin, K. V. and Saygin, M. Yu and Dyakonov, I. V. and Rakhlin, M. V. and Galimov, A. I. and Klimko, G. V. and Sorokin, S. V. and Sedova, I. V. and Kulagina, M. M. and Zadiranov, Yu M. and Toropov, A. A. and Evlashin, S. A. and Korneev, A. A. and Kulik, S. P. and Straupe, S. S.}, year={2025}, month=apr, pages={162–167},  }

@article{Son2018, title={High-efficiency broadband light coupling between optical fibers and photonic integrated circuits}, volume={7}, rights={De Gruyter expressly reserves the right to use all content for commercial text and data mining within the meaning of Section 44b of the German Copyright Act.}, ISSN={2192-8614}, DOI={10.1515/nanoph-2018-0075}, number={12}, journal={Nanophotonics}, publisher={De Gruyter}, author={Son, Gyeongho and Han, Seungjun and Park, Jongwoo and Kwon, Kyungmok and Yu, Kyoungsik}, year={2018}, month=dec, pages={1845–1864},  }

@misc{Soret2026,
      title={Quantum Energetic Advantage before Computational Advantage in Boson Sampling}, 
      author={Ariane Soret and Nessim Dridi and Stephen C. Wein and Valérian Giesz and Shane Mansfield and Pierre-Emmanuel Emeriau},
      year={2026},
      eprint={2601.08068},
      archivePrefix={arXiv},
      url={https://arxiv.org/abs/2601.08068}, 
}

@article{VanDerMeer2023, title={Experimental simulation of loop quantum gravity on a photonic chip}, volume={9}, rights={2023 The Author(s)}, ISSN={2056-6387}, DOI={10.1038/s41534-023-00702-y}, number={11}, journal={npj Quantum Information}, publisher={Nature Publishing Group}, author={van der Meer, Reinier and Huang, Zichang and Anguita, Malaquias Correa and Qu, Dongxue and Hooijschuur, Peter and Liu, Hongguang and Han, Muxin and Renema, Jelmer J. and Cohen, Lior}, year={2023}, month=apr, pages={1–7},  }

@article{Vigliar2021, title={Error-protected qubits in a silicon photonic chip}, volume={17}, rights={2021 The Author(s), under exclusive licence to Springer Nature Limited}, ISSN={1745-2481}, DOI={10.1038/s41567-021-01333-w}, number={10}, journal={Nature Physics}, publisher={Nature Publishing Group}, author={Vigliar, Caterina and Paesani, Stefano and Ding, Yunhong and Adcock, Jeremy C. and Wang, Jianwei and Morley-Short, Sam and Bacco, Davide and Oxenløwe, Leif K. and Thompson, Mark G. and Rarity, John G. and Laing, Anthony}, year={2021}, month=oct, pages={1137–1143},  }

@article{Wang2019, title={Boson Sampling with 20 Input Photons and a 60-Mode Interferometer in a $1{0}^{14}$-Dimensional Hilbert Space}, volume={123}, DOI={10.1103/PhysRevLett.123.250503}, number={25}, journal={Physical Review Letters}, publisher={American Physical Society}, author={Wang, Hui and Qin, Jian and Ding, Xing and Chen, Ming-Cheng and Chen, Si and You, Xiang and He, Yu-Ming and Jiang, Xiao and You, L. and Wang, Z. and Schneider, C. and Renema, Jelmer J. and Höfling, Sven and Lu, Chao-Yang and Pan, Jian-Wei}, year={2019}, month=dec, pages={250503} }

@article{Wang2020, title={Integrated photonic quantum technologies}, volume={14}, rights={2019 Springer Nature Limited}, ISSN={1749-4893}, DOI={10.1038/s41566-019-0532-1}, number={55}, journal={Nature Photonics}, publisher={Nature Publishing Group}, author={Wang, Jianwei and Sciarrino, Fabio and Laing, Anthony and Thompson, Mark G.}, year={2020}, month=may, pages={273–284},  }

@article{Heurtel2023percevalsoftware,
  doi = {10.22331/q-2023-02-21-931},
  url = {https://doi.org/10.22331/q-2023-02-21-931},
  title = {Perceval: {A} {S}oftware {P}latform for {D}iscrete {V}ariable {P}hotonic {Q}uantum {C}omputing},
  author = {Heurtel, Nicolas and Fyrillas, Andreas and Gliniasty, Gr{\'{e}}goire de and Le Bihan, Rapha{\"{e}}l and Malherbe, S{\'{e}}bastien and Pailhas, Marceau and Bertasi, Eric and Bourdoncle, Boris and Emeriau, Pierre-Emmanuel and Mezher, Rawad and Music, Luka and Belabas, Nadia and Valiron, Benoît and Senellart, Pascale and Mansfield, Shane and Senellart, Jean},
  journal = {{Quantum}},
  issn = {2521-327X},
  publisher = {{Verein zur F{\"{o}}rderung des Open Access Publizierens in den Quantenwissenschaften}},
  volume = {7},
  pages = {931},
  month = feb,
  year = {2023}
}

@article{phase_compilation,
  title = {High-fidelity quantum state evolution in imperfect photonic integrated circuits},
  author = {Mower, Jacob and Harris, Nicholas C. and Steinbrecher, Gregory R. and Lahini, Yoav and Englund, Dirk},
  journal = {Phys. Rev. A},
  volume = {92},
  pages = {032322},
  numpages = {7},
  year = {2015},
  month = {Sep},
  publisher = {American Physical Society},
  doi = {10.1103/PhysRevA.92.032322},
  url = {https://link.aps.org/doi/10.1103/PhysRevA.92.032322}
}

@INPROCEEDINGS{6804281,
  author={Verdú, Sergio},
  booktitle={2014 Information Theory and Applications Workshop (ITA)}, 
  title={Total variation distance and the distribution of relative information}, 
  year={2014},
  volume={},
  number={},
  pages={1-3},
  doi={10.1109/ITA.2014.6804281}}

@inproceedings{mqt,
  title        = {The {{MQT}} Handbook: {{A}} Summary of Design Automation Tools and Software for Quantum Computing},
  shorttitle   = {{The MQT Handbook}},
  author       = {Wille, Robert and Berent, Lucas and Forster, Tobias and Kunasaikaran, Jagatheesan and Mato, Kevin and Peham, Tom and Quetschlich, Nils and Rovara, Damian and Sander, Aaron and Schmid, Ludwig and Schoenberger, Daniel and Stade, Yannick and Burgholzer, Lukas},
  year         = 2024,
  booktitle    = qsw,
  doi          = {10.1109/QSW62656.2024.00013},
  eprint       = {2405.17543},
  eprinttype   = {arxiv},
  addendum     = {A live version of this document is available at \url{https://mqt.readthedocs.io}}
}

@INPROCEEDINGS{yannick,
  author={Stade, Yannick and Lin, Wan-Hsuan and Cong, Jason and Wille, Robert},
  booktitle=iccad, 
  title={Routing-Aware Placement for Zoned Neutral Atom-based Quantum Computing}, 
  year={2025},
  volume={},
  number={},
  pages={1-9},
  doi={10.1109/ICCAD66269.2025.11240721}}

@INPROCEEDINGS{11250295,
  author={Schoenberger, Daniel and Wille, Robert},
  booktitle=qce, 
  title={Orchestrating Multi-Zone Shuttling in Trapped-Ion Quantum Computers}, 
  year={2025},
  volume={01},
  number={},
  pages={1069-1075},
  doi={10.1109/QCE65121.2025.00119}}

@misc{zhu2025quantumcompilerdesignqubit,
      title={Quantum Compiler Design for Qubit Mapping and Routing: A Cross-Architectural Survey of Superconducting, Trapped-Ion, and Neutral Atom Systems}, 
      author={Chenghong Zhu and Xian Wu and Zhaohui Yang and Jingbo Wang and Anbang Wu and Shenggen Zheng and Xin Wang},
      year={2025},
      eprint={2505.16891},
      archivePrefix={arXiv},
      url={https://arxiv.org/abs/2505.16891}, 
}

@ARTICLE{10783033,
  author={Schoenberger, Daniel and Hillmich, Stefan and Brandl, Matthias and Wille, Robert},
  journal=tcad, 
  title={Shuttling for Scalable Trapped-Ion Quantum Computers}, 
  year={2025},
  volume={44},
  number={6},
  pages={2144-2155},
  doi={10.1109/TCAD.2024.3513262}}

@inproceedings{10.1145/3297858.3304075,
author = {Murali, Prakash and Baker, Jonathan M. and Javadi-Abhari, Ali and Chong, Frederic T. and Martonosi, Margaret},
title = {Noise-Adaptive Compiler Mappings for Noisy Intermediate-Scale Quantum Computers},
year = {2019},
isbn = {9781450362405},
url = {https://doi.org/10.1145/3297858.3304075},
doi = {10.1145/3297858.3304075},
booktitle = {Int'l Conf. on Architectural Support for Programming Languages and Operating Systems},
}

@inproceedings{10.1145/3297858.3304023,
author = {Li, Gushu and Ding, Yufei and Xie, Yuan},
title = {Tackling the Qubit Mapping Problem for NISQ-Era Quantum Devices},
year = {2019},
url = {https://doi.org/10.1145/3297858.3304023},
doi = {10.1145/3297858.3304023},
booktitle = {Int'l Conf. on Architectural Support for Programming Languages and Operating Systems},
pages = {1001–1014},
numpages = {14},
location = {Providence, RI, USA},
series = {ASPLOS '19}
}

@INPROCEEDINGS{10993241,
  author={Stade, Yannick and Schmid, Ludwig and Burgholzer, Lukas and Wille, Robert},
  booktitle=date, 
  title={Optimal State Preparation for Logical Arrays on Zoned Neutral Atom Quantum Computers}, 
  year={2025},
  volume={},
  number={},
  pages={1-7},
  doi={10.23919/DATE64628.2025.10993241}}

@article{Schmid_2024,
doi = {10.1088/2058-9565/ad33ac},
url = {https://doi.org/10.1088/2058-9565/ad33ac},
year = {2024},
month = {apr},
publisher = {IOP Publishing},
volume = {9},
number = {3},
pages = {033001},
author = {Schmid, Ludwig and Locher, David F and Rispler, Manuel and Blatt, Sebastian and Zeiher, Johannes and Müller, Markus and Wille, Robert},
title = {Computational capabilities and compiler development for neutral atom quantum processors—connecting tool developers and hardware experts},
journal = {Quantum Science and Technology}
}

\end{document}